\documentclass[american,prb,aps,amsmath,amssymb,reprint,superscriptaddress,showpacs]{revtex4-1}
\usepackage[T1]{fontenc}
\usepackage[utf8]{inputenc}
\usepackage{color}
\usepackage{babel}
\usepackage{textcomp}
\usepackage{amstext}
\usepackage{amssymb}
\usepackage{graphicx}
\usepackage[unicode=true,
 bookmarks=true,bookmarksnumbered=false,bookmarksopen=false,
 breaklinks=true,pdfborder={0 0 0},pdfborderstyle={},backref=false,colorlinks=true,pdfpagemode=FullScreen]
 {hyperref}
\hypersetup{pdftitle={Si(111) under tensile strain: evidence for the c(2x4) surface reconstruction},
 pdfauthor={R. Zhachuk, J. Coutinho, A. Dolbak, V. Cherepanov and B. Voigtländer},
 pdfsubject={68.35.bg, 68.35.Gy, 68.35.Md},
 pdfkeywords={Silicon, Germanium, Surface strain, Surface stress},
 allcolors=blue}
\usepackage{breakurl}



\begin{document}

\title{Si$(111)$ strained layers on Ge$(111)$: evidence for c$(2\times4)$
domains }

\author{R. Zhachuk}
\email{zhachuk@gmail.com}

\affiliation{Department of Physics \& I3N, University of Aveiro, Campus Santiago,
3810-193 Aveiro, Portugal}

\affiliation{Institute of Semiconductor Physics, pr. Lavrentyeva 13, Novosibirsk
630090, Russia}

\author{J. Coutinho}

\affiliation{Department of Physics \& I3N, University of Aveiro, Campus Santiago,
3810-193 Aveiro, Portugal}

\author{A. Dolbak}

\affiliation{Institute of Semiconductor Physics, pr. Lavrentyeva 13, Novosibirsk
630090, Russia}

\author{V. Cherepanov}

\affiliation{Peter Grünberg Institut (PGI-3), Forschungszentrum Jülich, 52425
Jülich, Germany}

\affiliation{JARA-Fundamentals of future Information Technology, 52455 Jülich,
Germany}

\author{B. Voigtländer}

\affiliation{Peter Grünberg Institut (PGI-3), Forschungszentrum Jülich, 52425
Jülich, Germany}

\affiliation{JARA-Fundamentals of future Information Technology, 52455 Jülich,
Germany}

\date{\today}
\begin{abstract}
The tensile strained Si$(111)$ layers grown on top of Ge$(111)$
substrates are studied by combining scanning tunneling microscopy,
low energy electron diffraction and first-principles calculations.
It is shown that the layers exhibit $c(2\times4)$ domains, which
are separated by domain walls along $\left\langle \bar{1}10\right\rangle $
directions. A model structure for the $c(2\times4)$ domains is proposed,
which shows low formation energy and good agreement with the experimental
data. The results of our calculations suggest that Ge atoms are likely
to replace Si atoms with dangling bonds on the surface (rest-atoms
and adatoms), thus significantly lowering the surface energy and inducing
the formation of domain walls. The experiments and calculations demonstrate
that when surface strain changes from compressive to tensile, the
(111) reconstruction converts from dimer-adatom-stacking fault-based
to adatom-based structures. 
\end{abstract}


\maketitle

\section{INTRODUCTION }

The importance of stress and strain fields on surface physics is well
recognized.\cite{mul04} They can have a strong impact on surface
reconstruction, stability of surface planes, step bunching and surface
diffusion.\cite{ter95,che02,che04,per12,zha13} The close chemistry
of Si and Ge, combined with a lattice mismatch of about 4\%, make
the Ge/Si system a prototypical model to study the effect of interfacial
elastic strain. Ge epitaxy on Si(111) has been extensively studied
and follows the Stranski-Krastanov growth mode.\cite{voi01} Here,
the formation of compressively strained Ge islands on Si(111) substrates
has attracted much interest due to their prospective use as template-structures
in nanoelectronics and nanophotonics.\cite{liu12,loz14}

First principles calculations of $(111)$ surface energies of silicon
and germanium have predicted a change of the surface structure when
the applied elastic strain changes from compressive to tensile.\cite{van87,mer93,zha13}
The surface reconstruction changes from dimer-adatom-stacking fault
(DAS) based to adatom-based according to the following sequence: $5\times5$
DAS (strongly compressive) $\rightarrow$ $7\times7$ DAS (weakly
compressive or relaxed) $\rightarrow$ adatom-based reconstructions
(relaxed or tensile). A few contenders for adatom-based reconstructions
of the (111) surfaces of Si and Ge have been proposed, all showing
close surface energy. These are $2\times2$, $c(2\times8)$, $c(2\times4)$,
and $\sqrt{3}\times\sqrt{3}$, and they were all observed experimentally
in Ge/Si$(111)$,\cite{koh91} with the first three being found on
quenched Si(111) surfaces as well.\cite{koi97} The adatom density
in the $2\times2$, $c(2\times8)$, $c(2\times4)$ reconstructions
is the same, and these structures differ only in their arrangement.
The density of adatoms in the $\sqrt{3}\times\sqrt{3}$ reconstruction
is $1/3$ higher. 

There are several experimental confirmations of the above-mentioned
sequence of structural changes. The fully relaxed Ge$(111)$ surface
adopts the adatom-based $c(2\times8)$ arrangement.\cite{bec89} Compressively
strained Ge layers and islands form during Ge molecular-beam epitaxy
(MBE) on Si$(111)$ substrates. In this case, Ge layers with compressive
biaxial strain above $\varepsilon\sim0.01$ have $5\times5$ DAS reconstruction,
while less strained layers show a $7\times7$ DAS structure.\cite{zha13}
On the other hand, for Si layers, compressively strained (111) terraces
on a stepped Si(111) surface show the tendency to form a $5\times5$
DAS reconstruction,\cite{kim10} while fully relaxed Si(111) adopts
the well-known $7\times7$ DAS reconstruction.\cite{tak85}

Despite the huge knowledge available for the Ge/Si system, important
issues remain to be addressed. For instance, structural data for tensile
strained Si$(111)$ are still missing. Filling in this gap is the
main goal of this work, and this is addressed by means of combining
first-principles atomistic modeling with scanning tunneling microscopy
(STM) and low energy electron diffraction (LEED) measurements of Si
layers grown on Ge$(111)$. We start by reporting the diffraction
data, then we move on to the STM data and finally, we describe the
results of the calculations, which provide insight to the experimental
results.

\section{METHODS}

\subsection{Experimental procedure}

The experiments were performed in two separate ultra-high vacuum systems,
one being equipped with a STM operating at room temperature in constant-current
mode and the second one with a LEED system. The STM chamber contains
the Si and Ge $e$-beam evaporators for deposition of Si and Ge by
MBE. A Si stripe heated with direct current was used as a source of
silicon atoms in the LEED chamber. A quartz crystal balance and STM
images were used to measure the deposited amount of Si and Ge. Si
was evaporated at a rate of 1 BL/min (BL stands for bilayer). The
samples were resistively heated with direct current. The temperature
of the substrate was measured using an infrared optical pyrometer. 

\begin{figure}
\includegraphics[clip,width=8.5cm]{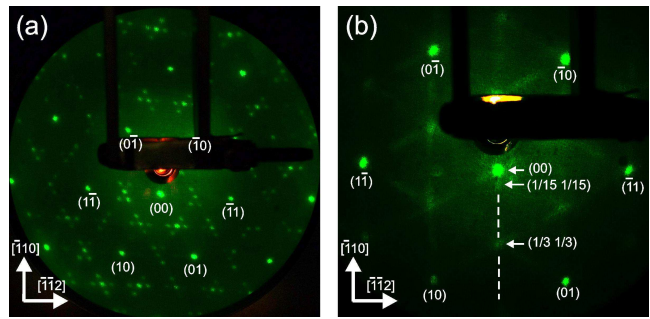}

\caption{\label{fig1}(color online). (a) LEED pattern from the clean Ge$(111)\textrm{-}c(2\times8)$
sample. Electron energy is $90$~eV. Integer order spots are marked.
See Figs.~1(a) and 1(b) in Supplemental Material at {[}\emph{URL
will be inserted by publisher}{]} for a schematics of the Ge$(111)\textrm{-}c(2\times8)$
diffraction pattern and marking of the fractional order spots. (b)
LEED pattern from the Ge$(111)$ surface after adsorption of the $\sim\!4$~BL
of silicon at $T_{\mathrm{ads}}=550\,\text{\textdegree}\mathrm{C}$.
Electron energy is $60$~eV. Integer order spots and spots at $(1/3\;1/3)$-like
and $(1/15\;1/15)$-like positions are marked. A faint streak along
the $[\bar{1}10]$ direction is highlighted by a dashed line.}
\end{figure}

The clean germanium surface was prepared in the LEED chamber by repeated
ion-bombardement cycles of Ar ($800$~eV), followed by annealing
at $800\,\mathrm{\text{\textdegree}C}$ of Ge$(111)$ samples until
a sharp $c(2\times8)$ diffraction pattern was observed (Fig.~\ref{fig1}(a)).
In the STM chamber, Ge$(111)$ was prepared by MBE growth of 3D relaxed
Ge islands on top of a clean Si$(111)$ surface. Formation details
of relaxed Ge islands on Si$(111)$ are given in Refs.~\onlinecite{che02,che04}.

\subsection{Computational details}

The surface energy (per unit area) of the reconstructed Si$(111)$
surface is defined and calculated as $\gamma_{\mathrm{rec}}(\varepsilon)=\gamma_{1\times1}(\varepsilon)-\triangle\gamma_{\mathrm{rec}}(\varepsilon)$,
following the procedure detailed in Ref.~\onlinecite{zha13}. Here,
$\gamma_{1\times1}(\varepsilon)$ is the energy of the unreconstructed
relaxed Si$(111)$-$1\!\times\!1$ surface as a function of applied
biaxial tensile strain $\varepsilon$, and $\triangle\gamma_{\mathrm{rec}}(\varepsilon)$
is a strain-dependent energy gain due to surface reconstruction. $\gamma_{1\times1}(\varepsilon)$
was calculated using a 12-Si-bilayer thick symmetric slab according
to the following expression,

\begin{equation}
\gamma_{1\times1}(\varepsilon)=\frac{1}{2S_{1\times1}(\varepsilon)}\left[E_{\mathrm{tot}}^{1\times1}(\varepsilon)-\mu(\varepsilon)N\right],\label{eq:1}
\end{equation}
where $\mu$ is the energy per Si atom in bulk under strain (Si chemical
potential), $S_{1\times1}$ is the area of a $1\times1$ surface cell
and $E_{\mathrm{tot}}^{1\times1}(\varepsilon)$ is the total energy
of the symmetric slab comprising \emph{N} atoms per simulation cell.
Two bilayers in the middle of the slab were kept frozen, while atoms
in other layers were allowed to move without any constraints during
atomic optimizations.

Each value of $\triangle\gamma_{\mathrm{rec}}(\varepsilon)$ was calculated
using two 6-bilayer thick slabs terminated by hydrogen on one side.
The first hydrogenated slab had an unreconstructed surface, while
the second had a surface reconstruction corresponding to the structure
under scrutiny. With this setup, the location of H and Si atoms at
the bottom layer was kept frozen during atomic optimizations, while
all other atoms were freely allowed to relax. The energy gain per
unit area upon reconstruction is therefore,

\begin{equation}
\triangle\gamma_{rec}(\varepsilon)=\frac{1}{S_{\mathrm{rec}}(\varepsilon)}\left[E_{\mathrm{tot}}^{\textrm{rec-H}}(\varepsilon)-E_{\mathrm{tot}}^{\textrm{1\ensuremath{\times}1-H}}(\varepsilon)M-\mu(\varepsilon)K\right],\label{eq:2}
\end{equation}
where $S_{\mathrm{rec}}$ is the unit cell area of the reconstructed
slab, $M=S_{\mathrm{rec}}/S_{1\times1}$ is the number of $1\!\times\!1$
reference cells spanned by a reconstructed cell, and \emph{K} accounts
for the number of Si surface atoms in excess to those in the reference
cell. In this expression $E_{\mathrm{tot}}^{\textrm{rec-H}}(\varepsilon)$
is the total energy of the reconstructed hydrogenated slab, while
$E_{\mathrm{tot}}^{\textrm{1\ensuremath{\times}1-H}}(\varepsilon)$
refers to the total energy of the unreconstructed hydrogenated slab
with a $1\!\times\!1$ surface cell.

Total energies were calculated from first principles by using the
density functional \textsc{siesta} code.\cite{sol02} The exchange-correlation
was treated within the local density approximation (LDA).\cite{per81}
Test calculations performed using the generalized gradient approximation
(GGA)\cite{per96} are reported in Table~1 of Supplemental Material
at {[}\emph{URL will be inserted by publisher}{]} confirming the suitability
of LDA to address our problem. The $\mathbf{k}$-space integrations
over Brillouin zones (BZ) were approximated by sums over Monkhorst-Pack
grids of $\mathbf{k}$-points.\cite{mon76} Norm-conserving pseudopotentials
were employed to account for electronic core states,\cite{tro91}
whereas valence states were represented by means of linear combinations
of numerical atomic orbitals of the Sankey-Niklewski type, generalized
to be arbitrarily complete with the inclusion of multiple $\zeta$
orbitals and polarization states.\cite{sol02} The calculations were
performed using double-$\zeta$ functions (DZP) basis for Si atoms
at the three upper slab layers and single-$\zeta$ functions (SZ)
for H as well as Si atoms at the remaining layers. Such choice for
the basis was previously shown to result in surface energies with
comparable accuracy to those using a full DZP basis.\cite{zha13}
Si atoms with DZP basis have two sets of \emph{s} and \emph{p} orbitals
plus one set of \emph{d} orbitals. Si atoms with SZ basis have one
set of \emph{s} and \emph{p} orbitals, while H atoms have a single
\emph{s} orbital.

The unreconstructed bottom surfaces were saturated by hydrogen atoms
making $1.50\,\mathrm{\mathring{A}}$ Si–H bonds. The opposite slab
surfaces were set up according to specific surface structure models.
These are $7\!\times\!7$ DAS,\cite{tak85} as well as single-domain
adatom-based $2\times2$, $c(2\times8)$, $c(2\times4)$, and $\sqrt{3}\times\sqrt{3}$
reconstructions.\cite{bec89,qin07} For the $c(2\times4)$ domains,
various widths and domain wall structure were considered. Adatoms
on $2\times2$, $c(2\times8)$, $c(2\times4)$, and $\sqrt{3}\times\sqrt{3}$
surfaces were placed at high-symmetry $T_{4}$ adsorption sites.

A uniform real-space grid equivalent to a plane-wave cutoff of $200\,\mathrm{Ry}$
was used for Fourier transforming the density and potential fields.
The geometry was optimized until all atomic forces became less than
$1$~meV/Å. Below this threshold, surface structures were considered
to have attained equilibrium. All periodic slabs were separated by
a $30$~Å thick vacuum layer. Under these conditions, converged calculations
using a bulk conventional unit cell with a $8\times8\times8$ BZ-sampling
grid gave a lattice constant of Si $a_{\mathrm{Si}}=5.420$~Å. We
used specific $\mathbf{k}$-point grids for each surface reconstruction/slab,
depending on its respective lateral dimensions, namely: $20\times20\times1$
for $1\times1$, $3\times3\times1$ for $7\times7$ DAS, $10\times10\times1$
for $2\times2$, $8\times2\times1$ for $c(2\times8)$ (rectangular
surface cell), $12\times12\times1$ for $\sqrt{3}\times\sqrt{3}$,
and $10\times12\times1$ for single-domain $c(2\times4)$. The $\mathbf{k}$-point
grids for $c(2\times4)$ with variable domains were dependent on the
domain width: $5\times12\times1$ for domains comprising 1 adatom
in width, $4\times12\times1$ for domains comprising 2 adatoms, $3\times12\times1$
for domains comprising 3 adatoms, $2\times12\times1$ for domains
comprising 4-6 adatoms in width. The resulting $\mathbf{k}$-point
surface densities in reciprocal space are approximately the same for
all cells. Convergence tests regarding the BZ sampling, slab type
and thickness, as well as the basis functions, were reported elsewhere.\cite{zha13} 

\begin{figure}
\includegraphics[clip,width=8cm]{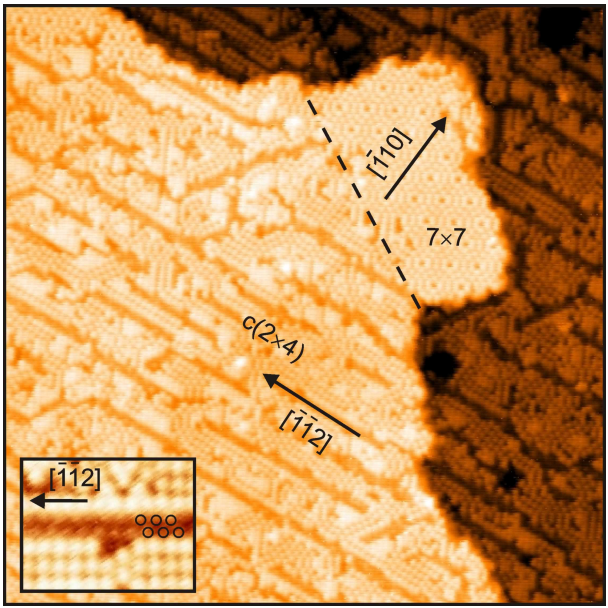}

\caption{\label{fig2}(color online). STM image of the MBE-grown $2$~BL silicon
on top of a relaxed Ge$(111)$ island, $T_{\mathrm{ads}}=540\,\text{\textdegree}\mathrm{C}$.
The image exhibits the $7\times7$ area and somewhat disordered $c(2\times4)$
domains. The dashed line highlights the domain boundary between $7\times7$
and $c(2\times4)$ surfaces. Image size is $640\times640\,\mathrm{\mathring{A}^{2}}$.
$U=+1.8$~V and $I=1.0$~nA. The inset shows a high resolution STM
image of the DW. The internal zig-zag structure of the DW is highlighted
by black circles. The inset dimensions are $71\times54\,\mathrm{\mathring{A}^{2}}$.}
\end{figure}

The constant-current STM images were produced within the Tersoff-Hamann
approach.\cite{ter85} The \textsc{WSXM} software was used to process
the experimental and calculated STM images.\cite{hor07} 

\section{RESULTS AND DISCUSSION}

\subsection{Experimental STM and LEED results}

After cleaning the germanium samples the surface exhibits the well-known
$c(2\times8)$ diffraction pattern, typical for the clean relaxed
Ge$(111)$ surface (see Fig.~\ref{fig1}(a)).\cite{raz09} The LEED
pattern after deposition of $4$~BL of silicon on Ge$(111)$ surface
at $T_{\mathrm{ads}}=550\,\text{\textdegree}\mathrm{C}$ is shown
in Figure~\ref{fig1}(b). Here the spots from the $c(2\times8)$
surface reconstruction are completely vanished, and instead, the diffraction
pattern shows blurred spots at $(1/3\;1/3)$-like positions, faint
streaks along $\left\langle \bar{1}10\right\rangle $ directions and
weak spots at about $(1/15\;1/15)$-like positions close to the $(0\;0)$
central spot. Similar diffraction patterns were observed for the coverage
range $\Theta_{\mathrm{Si}}=2\textrm{-}4$~BL and in the temperature
range $T_{\mathrm{ads}}=400\textrm{-}550\,\text{\textdegree}\mathrm{C}$.
The appearance of the spots at the $(1/3\;1/3)$-like positions after
Si MBE growth on Ge$(111)$ surface was also reported in Ref.~\onlinecite{ich84}.

\begin{figure}
\includegraphics[clip,width=8.5cm]{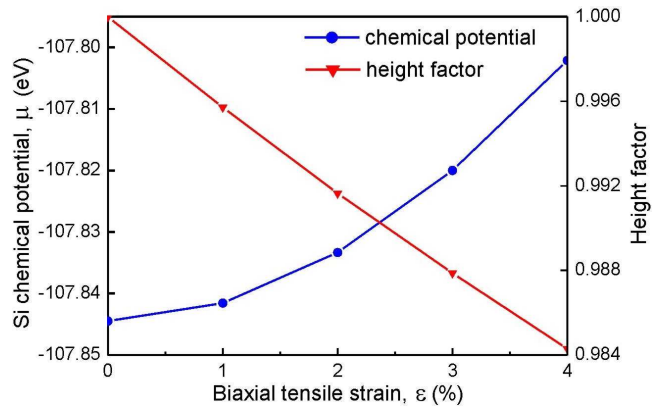}

\caption{\label{fig3}(color online). (left y-axis) Si chemical potential and
(right y-axis) height factor, $f_{h}$ (see text) for Si$(111)$ interplanar
spacing, calculated for different biaxial strain states $\varepsilon$
in the $(111)$ plane.}
\end{figure}

An STM image of $2$~BL silicon deposited on top of a relaxed 3D
Ge island at $T_{\mathrm{ads}}=540\,\text{\textdegree}\mathrm{C}$
is shown in Fig.~\ref{fig2}. The image shows two $(111)$ terraces
separated by a step. The structure on the terraces shows significant
disorder beyond the nanometer scale, but at a smaller scale we can
clearly distinguish the existence of predominant patterns. Such observation
strongly indicates the existence of several surface reconstructions
with close formation energies. Most part of the surface is covered
with a reconstruction having a rectangular unit cell. This structure
looks very similar to the $c(2\times4)$ reconstruction found in quenched
Si$(111)$ surfaces.\cite{koi97} The “peninsula” towards the upper-right
corner of Figure~\ref{fig2} exhibits a $7\times7$ surface reconstruction,
typical of clean relaxed Si$(111)$ surfaces.

The plain $c(2\times4)$ reconstruction has a rectangular cell and
can form in three rotational domains each rotated by $120\text{\textdegree}$
as follows from the three-fold $C_{3v}$ symmetry of the $(111)$
substrate. The actual surface structure in Fig.~\ref{fig2} consists
of local patches of $c(2\times4)$ reconstruction separated by domain
walls (DWs) and oriented along the same direction at small scale.
The DWs are the dark stripes in Fig.~\ref{fig2} representing surface
depressions or trenches. Since the atomic structure of the Si$(111)$-7$\times$7
reconstruction is known,\cite{tak85} the crystallographic directions
in the STM image in Fig.~\ref{fig2} are readily obtained. Thus,
it was found that the DWs are elongated along $\left\langle \bar{1}\bar{1}2\right\rangle $
directions. The shorter side of the $c(2\times4)$ unit cell is $\sqrt{3}a$
long, where $a$ is the unit length of the unreconstructed $(111)$
surface, and it is parallel to the DW directions. Conversely, the
longer side of the $c(2\times4)$ unit cell (which is $2a$ long)
is perpendicular to the DWs (along $\left\langle \bar{1}10\right\rangle $
directions). The typical $c(2\times4)$ domain in Fig.~\ref{fig2}
consists of 3 rows of bright spots (Si adatoms) along $\left\langle \bar{1}10\right\rangle $.
The inset in Fig.~\ref{fig2} shows a high-resolution STM image of
two $c(2\times4)$ patches separated by a DW. The internal structure
of the DW is resolved, exhibiting a zig-zag row of dimmed spots highlighted
by the small black circles. 

\subsection{Theoretical results and comparison with experimental data}

\begin{figure}
\includegraphics[width=6.5cm]{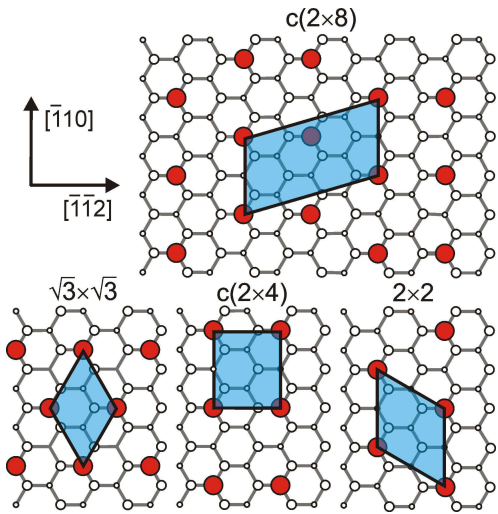}\caption{\label{fig4}(color online). Schematics of adatom-based Si(111) surface
reconstructions: $c(2\times8)$, $\sqrt{3}\times\sqrt{3}$, $c(2\times4)$,
and $2\times2$. A single silicon bilayer with adsorbed silicon atoms
is shown. Big and small white-filled circles represent upper and lower
atoms of the bilayer, respectively. Large red-filled circles represent
Si adatoms. The unit cell for each reconstruction is also outlined.}
\end{figure}

Figure~\ref{fig3} shows the calculated silicon chemical potential
as a function of $(111)$ biaxial strain in the bulk. From elasticity,
it follows that a $(111)$-biaxially strained cubic solid leads to
an opposite strain along $[111]$. Such effect has to be accounted
for in strained surface calculations, and this is done by letting
the surfaces to freely relax towards the vacuum. In bulk, this effect
was considered by using appropriately strained $1\times1$ bulk-slabs
with several heights, $h$, related to the strain-free height $h_{0}$
by a height factor, $f_{h}=h/h_{0}$. We determined their equilibrium
heights ($h$ values that minimized the energy), which were then used
to obtain the energy per Si atom in bulk under strain. Figure~\ref{fig3}
also depicts the calculated height factors for Si $1\times1$ bulk-slabs.
It shows how the equilibrium distance between $(111)$ layers in bulk
depends on the applied biaxial strain.

Several $(111)$ adatom-based reconstructions were considered in this
study: $2\times2$, $c(2\times8)$, $c(2\times4)$, and $\sqrt{3}\times\sqrt{3}$
(Fig.~\ref{fig4}). These are the lowest energy configurations, and
therefore the most probable adatom-based $(111)$ reconstructions
of Si and Ge. The $2\times2$, $c(2\times8)$ and $c(2\times4)$ reconstructions
were observed on clean quenched surfaces of Si$(111)$ and at Ge/Si$(111)$
growth \cite{koi97,koh91}. The formation of the $\sqrt{3}\times\sqrt{3}$
surface reconstruction was observed on MBE-grown Ge/Si$(111)$ as
reported in Refs.~\onlinecite{qin07,zha13}.

\begin{figure}
\includegraphics[width=7.5cm]{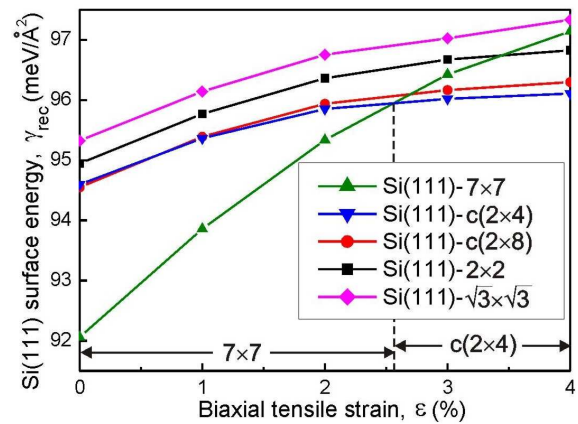}

\caption{\label{fig5}Si$(111)$ surface formation energies calculated for
tensile biaxial strain states in the range $\varepsilon=0\%\textrm{-}4\%$
for $7\times7$, $c(2\times8)$, $\sqrt{3}\times\sqrt{3}$, $c(2\times4)$,
$2\times2$ surface reconstructions.}
\end{figure}

Figure~\ref{fig5} shows the calculated surface energies $\gamma_{\mathrm{rec}}$
of Si$(111)$ as a function of applied tensile strain $\varepsilon$
for various experimentally observed reconstructions. In agreement
with previous studies, we find that adatom-based structures are more
stable than $7\times7$ DAS-based structure when biaxial tensile strain
above $\sim\!2.5\%$ is applied to Si$(111)$.\cite{zha13,van87,mer93}
As opposed to the simple adatom-based reconstructions, the presence
of several reconstruction elements (dimers, adatoms and stacking faults)
in DAS-based surfaces make these intrinsically compressive.\cite{zha13}
Hence, with increasing the surface tensile strain, the surface formation
energy of DAS-based reconstructions grows faster than that of adatom-based
structures. As one can see from Fig.~\ref{fig5}, the lowest energy
reconstruction at $\varepsilon\gtrsim2.5\%$ is $c(2\times4)$. This
result explains the STM observations depicted in Fig.~\ref{fig2}.

\begin{figure}
\includegraphics[width=6.5cm]{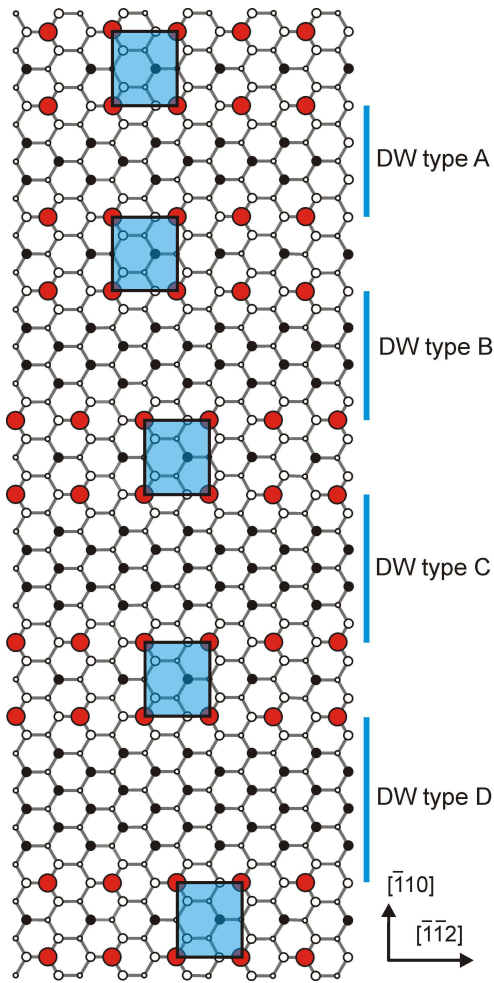}

\caption{\label{fig6}Schematics of possible DW atomic structures, from type
A to type D. Big white-filled circles represent upper atoms of the
bilayer with all bonds saturated; Black-filled circles represent upper
atoms of the bilayer with one dangling bond (rest-atoms); Small white-filled
circles represent lower atoms of the bilayer; Large red-filled circles
represent Si adatoms. The $c(2\times4)$ unit cell is outlined on
each domain.}
\end{figure}

Four contenders for the DW atomic structures that separate neighboring
$c(2\times4)$ domains were investigated and are represented in Fig.~\ref{fig6}.
The proposed DW models are simply bare $(111)$ substrate areas with
increasing width (types A-D). The width of these DWs is in approximate
agreement with the experimental STM image in Fig.~\ref{fig2}. For
the sake of space saving, the width of the $c(2\times4)$ domains
as represented in Fig.~\ref{fig6} is limited to two Si atomic rows.
The surface combining the Si$(111)\textrm{-}c(2\times4)$ domains
with DWs was simulated using periodic boundary conditions, keeping
the domain width fixed, but considering variable-width DW models (as
shown in Fig.~\ref{fig6}).

\begin{figure}
\includegraphics[width=7.5cm]{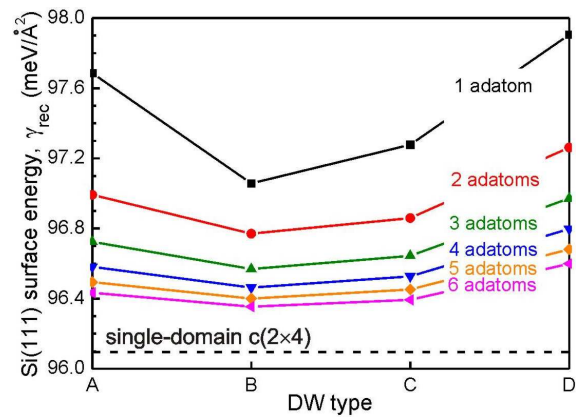}

\caption{\label{fig7}Si$(111)\textrm{-}c(2\times4)$ surface formation energies
calculated for 4\% tensile biaxial strain, as a function of the DW
width (corresponding to four different $c(2\times4)$ DW structures,
A-D). Each plot correspond to a specific $c(2\times4)$ domain width
measured in number of adatoms between neighboring DWs.}
\end{figure}

\begin{figure}
\includegraphics[width=8.5cm]{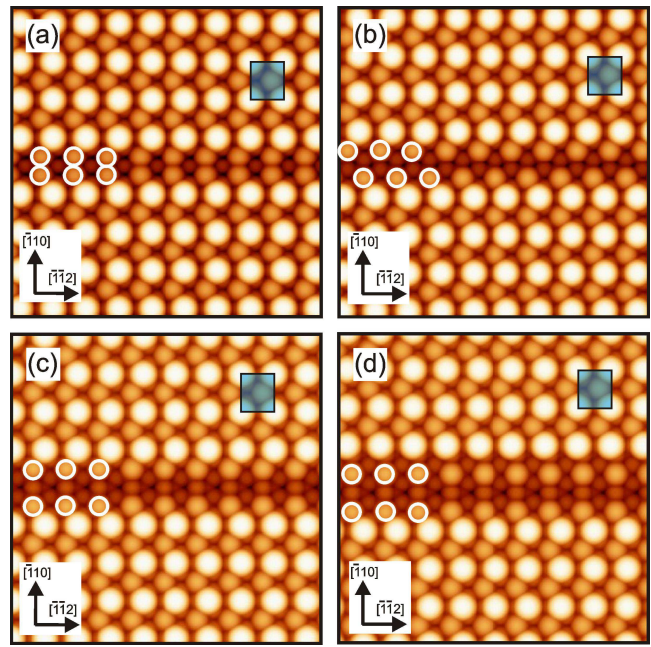}

\caption{\label{fig8}Calculated STM images of four different types of DWs
structures: (a) DW type A. (b) DW type B. (c) DW type C. (d) DW type
D. Bias voltage corresponds to $+1.0$~eV with respect to the theoretical
Fermi level. The $c(2\times4)$ unit cells are outlined. The structure
of DWs is highlighted by white circles.}
\end{figure}

Figure~\ref{fig7} shows that irrespectively of the domain width,
the Si$(111)\textrm{-}c(2\times4)$ surface with DW type B shows lowest
formation energy and therefore is expected to correspond to the observations.
The calculated STM images of the Si$(111)\textrm{-}c(2\times4)$ surface
near the DW area (DW types A-D) are presented in Figs.~\ref{fig8}(a)-\ref{fig8}(d),
respectively. The DW internal structure exhibits relatively bright
spots, somewhat dimmed as compared to those of Si adatoms. Some of
these spots are highlighted by white circles. Similar spots are also
observed within the $c(2\times4)$ domains, between Si adatoms. All
these spots are due to substrate rest-atoms raised after slab relaxation.
As shown in Fig.~\ref{fig8}(b), only DW type B is compatible with
the experimentally observed zig-zag structure as depicted in the inset
of Fig.~\ref{fig2}. This corroborates the assignment of the DW structure
to type B based on its low surface energy. It is noteworthy to mention
that DW type D (shown in Fig.~\ref{fig6}), spanning 6 parallel rows
of rest-atoms, is \emph{a priori} expected to show a zig-zag pattern
in STM due to rows 1 and 6 edging the $c(2\times4)$ domains. Instead,
we found that in the ground-state, spots from parallel rows 2 and
6 (or 1 and 5) are “switched on”, adding further support to our assignment
of type B structure to the observed DWs.

We can conjecture that the reason behind the appearance of DWs is
the partial relaxation of the silicon layers under tensile strain.
Accordingly, the observed domain and DW widths must be governed by
a balance between the energy gain from surface strain relief and the
energy penalty due to formation of unsaturated bonds across the bare
surface area of the DWs. Figure~\ref{fig2} suggests that the crossing
point involving these two factors takes place for a domain width equivalent
to about 3 adatoms. Still, the observed surface structure should correspond
to an energy state below that of single-domain $c(2\times4)$ reconstruction
(without DWs), and that is not what Fig.~\ref{fig7} shows. In fact,
the calculated data suggest that DWs should not appear in strained
Si$(111)$ under thermodynamic equilibrium conditions. Despite having
been grown at relatively high temperatures, and therefore likely to
be close to thermodynamic equilibrium, Fig.~\ref{fig7} shows that
domains with 3 adatoms width are metastable by $\sim\!0.5$~meV/Å$^{2}$
(with respect to a full covered $c(2\times4)$ surface). We also note
that the $c(2\times4)$ domains were observed in rather thin ($2\textrm{-}4$~BL
thick) Si layers, so that the number of atomic layers used in calculations
should be appropriate to describe the strain within the slab.

Until now, we have missed an ingredient in calculations which can,
in principle, play a relevant role in DW formation. That is Si/Ge
intermixing. Unfortunately, the chemical resemblance between Si and
Ge prevents their discrimination on the surface by using STM imagery
like that shown in Fig.~\ref{fig2}. Surface termination with bismuth
allows to distinguish between Si and Ge atoms,\cite{mys10,zha11}
but that would alter the surface reconstruction. The $(111)\textrm{-}7\times7$
reconstructed ``peninsula'' in Fig.~\ref{fig2} may suggest that
the surface consists of clean silicon layers, where strain is released
through the surrounding step edges. Clean Ge$(111)$ layers, which
can also form $7\times7$ reconstruction under compressive strain,\cite{zha13}
can be ruled out due to the absence of compressive strain in the surface.
Yet, we cannot exclude the formation of a Si-rich SiGe$(111)\textrm{-}7\times7$
reconstruction. 

In order to understand how Si/Ge intermixing can change the surface
energy and its structure, we performed a set of exploratory calculations.
Accordingly, we used Si/Ge$(111)$ hydrogenated slabs, where both
Si and Ge atoms had DZP basis set. The first (reference) slab had
$3$~BL of pure Si on top of $3$~BL of pure Ge. In the second slab,
we kept the same Si/Ge layered structure, but one atom at the topmost
Ge layer (at the Si/Ge interface) was replaced by Si, while one Si
atom at the surface was replaced by Ge. The calculated energy differences
show that the exchange of Si and Ge atoms leads to an energy drop
of up to $0.4$~eV per atom pair. We suggest that most energy gain
comes from the less reactive Ge dangling bond (when compared to that
of Si) when Ge occupies rest-atom or adatom sites, both at the $c(2\times4)$
domain and at the DW surfaces. Surface sites with saturated bonds
show less energy gain or no gain at all. The structural models used
to calculate Si/Ge intermixing and corresponding surface energies
are summarized in Table~2 of Supplemental Material at {[}\emph{URL
will be inserted by publisher}{]}.

Next, we constructed a $c(2\times4)$ slab with domain width equivalent
to 3 adatoms and DW type B, where \emph{all} atoms with dangling bonds
at the Si surface were replaced by Ge. Concurrently, an equal number
of Ge atoms at the topmost layer of Ge (at the Si/Ge interface) was
replaced by Si. This Ge-terminated surface shows an energy gain of
$14.2$~meV/Å$^{2}$ with respect to the reference slab (with no
Si/Ge intermixing). The single-domain $c(2\times4)$ reconstruction
without DWs shows a 1~meV/Å$^{2}$ lower energy gain. This difference
is naturally explained by the two times higher density of dangling
bonds in the DW area as compared to that in the $c(2\times4)$ domains.
Hence, the impact of Si/Ge intermixing on the surface energy is more
pronounced in the DW area. Combining surface energies calculated for
strained Si$(111)$ layers (Fig.~\ref{fig7}) with energy gains due
to Si/Ge intermixing, we estimate surface energies of single-domain
$c(2\times4)$ and $c(2\times4)$ with DWs (3 adatoms width) as $82.9$~meV/Å$^{2}$
and $82.4$~meV/Å$^{2}$, respectively. This makes the formation
of DWs energetically favorable when Si/Ge intermixing is taken into
account. Auger electron spectroscopy or photoelectron spectroscopy
measurements could help to determine if silicon layers are actually
terminated by Ge atoms.

\begin{figure}
\includegraphics[width=8.5cm]{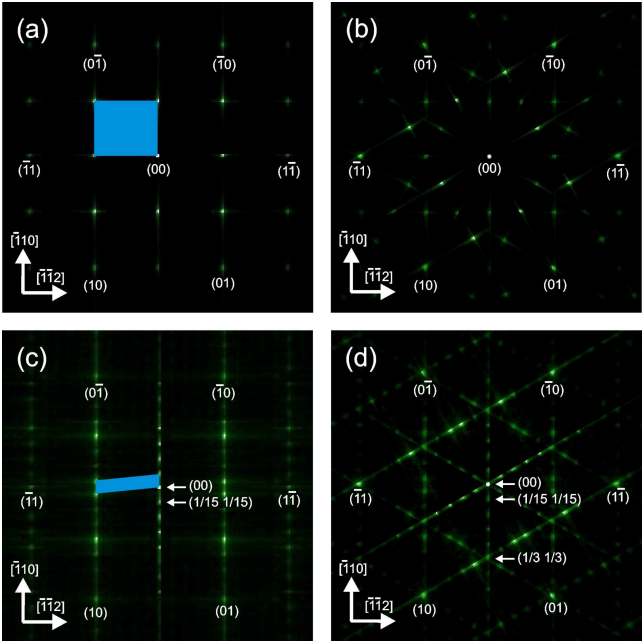}

\caption{\label{fig9}Calculated LEED patterns. (a) single-domain $c(2\times4)$
reconstruction (no DWs). (b) Three rotational domains (no DWs). (c)
$c(2\times4)$ domains with 3 adatoms width (one rotational domain)
separated by DWs type B. (d) $c(2\times4)$ domains with 3 adatoms
width (three rotational domains) separated by DWs type B. Reciprocal
unit cells are outlined in (a) and (c). See Figs. 2(a) and (b) of
Supplemental Material at {[}\emph{URL will be inserted by publisher}{]}
for the simulated STM images used to compute the LEED patterns in
(a) and (c).}
\end{figure}

The calculated LEED patterns representing fast Fourier transforms
of calculated STM images, are shown in Fig.~\ref{fig9}. Fig.~\ref{fig9}(a)
shows the calculated diffraction pattern from single-domain $c(2\times4)$
reconstruction without DWs, where only one rotational domain is present.
Fig.~\ref{fig9}(b) shows a diffraction pattern from the three possible
rotational domains of $c(2\times4)$ reconstruction (still without
DWs). The calculated LEED patterns in Figs.~\ref{fig9}(a) and \ref{fig9}(b)
cannot account for the experimental LEED pattern shown in \ref{fig1}(b).
We clearly have to consider the effect of DWs on the LEED patterns.

Fig.~\ref{fig9}(c) shows the calculated diffraction pattern from
one rotational domain of $c(2\times4)$ reconstruction with type-B
DWs. A domain width equivalent to 3 adatoms was considered. The translational
unit cell for the surface with DWs includes few $c(2\times4)$ cells
and is not rectangular (see Fig. 2(b) of Supplemental Material at
{[}\emph{URL will be inserted by publisher}{]}). One can see that
the spots from the $c(2\times4)$ reconstruction are now split along
the $[\bar{1}10]$ direction (perpendicular to the DW orientation).
This is the effect of intensity modulation by DWs. The resulting reciprocal
unit cell is smaller than that of the $c(2\times4)$ reconstruction
without DWs, and consequently it is not rectangular as well. The influence
of surface defects (including DWs) on LEED patterns was extensively
studied in the past (see for instance Ref.~\onlinecite{hen82} and
references therein). The split size ($\delta$) is inversely proportional
to the periodicity of $c(2\times4)$ domains ($\Gamma$) in the direction
perpendicular to DWs:

\begin{equation}
\Gamma=w_{\mathrm{D}}+w_{\mathrm{DW}}=\frac{100\%}{\delta\%}\times a_{\mathrm{Ge}(111)}\sin60\text{\textdegree},\label{eq:3}
\end{equation}
where 100\% of the surface BZ corresponds to the distance between
the $(0\;0)$ spot and integer first-order spots, while $a_{\mathrm{Ge}(111)}$
is the lattice constant of the unreconstructed Ge$(111)\textrm{-}1\times1$
surface. $w_{\mathrm{D}}$ and $w_{\mathrm{DW}}$ stand for domain
and DW widths, respectively. Thus, combining Eq.~\ref{eq:3} with
the experimental LEED pattern in Fig.~\ref{fig1}(b) we may estimate
the \emph{macroscopic} average domain width.\cite{hen82}

Fig.~\ref{fig9}(d) shows the calculated diffraction pattern from
all three rotational domains of $c(2\times4)$ reconstruction with
DW type B. The pattern shows many similarities with the experimental
LEED results shown in Fig.~\ref{fig1}(b). The split size in Fig.~\ref{fig9}(d)
is about 7\% which is equivalent to the $(1/15\;1/15)$-like spots
visible in Fig.~\ref{fig1}(b). From the LEED data, we obtain an
average $c(2\times4)$ domain width equivalent to 3 adatoms, which
fits nicely to the STM results. All six $(1/15\;1/15)$-like spots
with similar intensity are visible around the $(0\;0)$-spot in Fig.~\ref{fig1}(b).
Therefore, each $c(2\times4)$ rotational domain occupies similar
surface areas. Only first-order $(1/15\;1/15)$ fractional spots are
visible in experimental LEED pattern in Fig.~\ref{fig1}(b), while
higher-order spots are smeared out and become weak streaks along $\left\langle \bar{1}10\right\rangle $-like
directions. The intersections of these streaks give rise to $(1/3\;1/3)$-like
spots with weak intensity. The suppression of higher order fractional
spots in diffraction patterns is due to irregular widths of $c(2\times4)$
domains. A similar effect was observed, for example, on LEED patterns
of nickel contaminated Si$(100)$ containing irregular $2\times1$
domains.\cite{dol89,ned96}

\section{CONCLUSIONS}

Tensile strained Si$(111)$ prepared by silicon MBE growth on Ge$(111)$
substrates and on relaxed Ge$(111)$ template structures on top of
3D islands, was studied by low energy electron diffraction, scanning
tunneling microscopy and first-principles calculations. We show that
the calculated Si(111) surface reconstructions and their respective
energy ordering as a function of strain, match the experimental observations.
Namely, it is shown that under tensile strain the Si$(111)$ surface
exhibits domains of adatom-based $c(2\times4)$ reconstruction, separated
by domain walls. This contrasts with the relaxed and compressive strain
regimes where dimer-adatom-stacking fault structures are seen. An
atomic model for the domain wall that separates neighboring $c(2\times4)$
domains is proposed, showing low surface energy and good agreement
with the experimental microscopy and diffraction data. The average
domain width is also reported. According to the calculations, the
formation of domain walls on pure Si$(111)$ surface always imply
an energy penalty, suggesting that their appearance is unfavorable
under thermodynamic equilibrium conditions. We suggest that Ge/Si
intermixing can stabilize the DWs, hence explaining this apparent
contradiction.
\begin{acknowledgments}
We would like to thank the Novosibirsk State University for providing
the computational resources. This work was funded by the Fundação
para a Ciência e a Tecnologia (FCT) under the contract UID/CTM/50025/2013,
and by FEDER funds through the COMPETE 2020 Program.
\end{acknowledgments}

\bibliographystyle{apsrev4-1}

\begin{thebibliography}{31}%
\makeatletter
\providecommand \@ifxundefined [1]{%
 \@ifx{#1\undefined}
}%
\providecommand \@ifnum [1]{%
 \ifnum #1\expandafter \@firstoftwo
 \else \expandafter \@secondoftwo
 \fi
}%
\providecommand \@ifx [1]{%
 \ifx #1\expandafter \@firstoftwo
 \else \expandafter \@secondoftwo
 \fi
}%
\providecommand \natexlab [1]{#1}%
\providecommand \enquote  [1]{``#1''}%
\providecommand \bibnamefont  [1]{#1}%
\providecommand \bibfnamefont [1]{#1}%
\providecommand \citenamefont [1]{#1}%
\providecommand \href@noop [0]{\@secondoftwo}%
\providecommand \href [0]{\begingroup \@sanitize@url \@href}%
\providecommand \@href[1]{\@@startlink{#1}\@@href}%
\providecommand \@@href[1]{\endgroup#1\@@endlink}%
\providecommand \@sanitize@url [0]{\catcode `\\12\catcode `\$12\catcode
  `\&12\catcode `\#12\catcode `\^12\catcode `\_12\catcode `\%12\relax}%
\providecommand \@@startlink[1]{}%
\providecommand \@@endlink[0]{}%
\providecommand \url  [0]{\begingroup\@sanitize@url \@url }%
\providecommand \@url [1]{\endgroup\@href {#1}{\urlprefix }}%
\providecommand \urlprefix  [0]{URL }%
\providecommand \Eprint [0]{\href }%
\providecommand \doibase [0]{http://dx.doi.org/}%
\providecommand \selectlanguage [0]{\@gobble}%
\providecommand \bibinfo  [0]{\@secondoftwo}%
\providecommand \bibfield  [0]{\@secondoftwo}%
\providecommand \translation [1]{[#1]}%
\providecommand \BibitemOpen [0]{}%
\providecommand \bibitemStop [0]{}%
\providecommand \bibitemNoStop [0]{.\EOS\space}%
\providecommand \EOS [0]{\spacefactor3000\relax}%
\providecommand \BibitemShut  [1]{\csname bibitem#1\endcsname}%
\let\auto@bib@innerbib\@empty
\bibitem [{\citenamefont {M{\"u}ller}\ and\ \citenamefont
  {Sa{\'u}l}(2004)}]{mul04}%
  \BibitemOpen
  \bibfield  {author} {\bibinfo {author} {\bibfnamefont {P.}~\bibnamefont
  {M{\"u}ller}}\ and\ \bibinfo {author} {\bibfnamefont {A.}~\bibnamefont
  {Sa{\'u}l}},\ }\href {\doibase 10.1016/j.surfrep.2004.05.001} {\bibfield
  {journal} {\bibinfo  {journal} {Surf. Sci. Rep.}\ }\textbf {\bibinfo {volume}
  {54}},\ \bibinfo {pages} {157} (\bibinfo {year} {2004})}\BibitemShut
  {NoStop}%
\bibitem [{\citenamefont {Tersoff}\ \emph {et~al.}(1995)\citenamefont
  {Tersoff}, \citenamefont {Phang}, \citenamefont {Zhang},\ and\ \citenamefont
  {Lagally}}]{ter95}%
  \BibitemOpen
  \bibfield  {author} {\bibinfo {author} {\bibfnamefont {J.}~\bibnamefont
  {Tersoff}}, \bibinfo {author} {\bibfnamefont {Y.~H.}\ \bibnamefont {Phang}},
  \bibinfo {author} {\bibfnamefont {Z.}~\bibnamefont {Zhang}}, \ and\ \bibinfo
  {author} {\bibfnamefont {M.~G.}\ \bibnamefont {Lagally}},\ }\href {\doibase
  10.1103/PhysRevLett.75.2730} {\bibfield  {journal} {\bibinfo  {journal}
  {Phys. Rev. Lett.}\ }\textbf {\bibinfo {volume} {75}},\ \bibinfo {pages}
  {2730} (\bibinfo {year} {1995})}\BibitemShut {NoStop}%
\bibitem [{\citenamefont {Cherepanov}\ and\ \citenamefont
  {Voigtl{\"a}nder}(2002)}]{che02}%
  \BibitemOpen
  \bibfield  {author} {\bibinfo {author} {\bibfnamefont {V.}~\bibnamefont
  {Cherepanov}}\ and\ \bibinfo {author} {\bibfnamefont {B.}~\bibnamefont
  {Voigtl{\"a}nder}},\ }\href {\doibase 10.1063/1.1530730} {\bibfield
  {journal} {\bibinfo  {journal} {Appl. Phys. Lett.}\ }\textbf {\bibinfo
  {volume} {81}},\ \bibinfo {pages} {4745} (\bibinfo {year}
  {2002})}\BibitemShut {NoStop}%
\bibitem [{\citenamefont {Cherepanov}\ and\ \citenamefont
  {Voigtl{\"a}nder}(2004)}]{che04}%
  \BibitemOpen
  \bibfield  {author} {\bibinfo {author} {\bibfnamefont {V.}~\bibnamefont
  {Cherepanov}}\ and\ \bibinfo {author} {\bibfnamefont {B.}~\bibnamefont
  {Voigtl{\"a}nder}},\ }\href {\doibase 10.1103/PhysRevB.69.125331} {\bibfield
  {journal} {\bibinfo  {journal} {Phys. Rev. B}\ }\textbf {\bibinfo {volume}
  {69}},\ \bibinfo {pages} {125331} (\bibinfo {year} {2004})}\BibitemShut
  {NoStop}%
\bibitem [{\citenamefont {Persichetti}\ \emph {et~al.}(2012)\citenamefont
  {Persichetti}, \citenamefont {Sgarlata}, \citenamefont {Mattoni},
  \citenamefont {Fanfoni},\ and\ \citenamefont {Balzarotti}}]{per12}%
  \BibitemOpen
  \bibfield  {author} {\bibinfo {author} {\bibfnamefont {L.}~\bibnamefont
  {Persichetti}}, \bibinfo {author} {\bibfnamefont {A.}~\bibnamefont
  {Sgarlata}}, \bibinfo {author} {\bibfnamefont {G.}~\bibnamefont {Mattoni}},
  \bibinfo {author} {\bibfnamefont {M.}~\bibnamefont {Fanfoni}}, \ and\
  \bibinfo {author} {\bibfnamefont {A.}~\bibnamefont {Balzarotti}},\ }\href
  {\doibase 10.1103/PhysRevB.85.195314} {\bibfield  {journal} {\bibinfo
  {journal} {Phys. Rev. B}\ }\textbf {\bibinfo {volume} {85}},\ \bibinfo
  {pages} {195314} (\bibinfo {year} {2012})}\BibitemShut {NoStop}%
\bibitem [{\citenamefont {Zhachuk}\ \emph {et~al.}(2013)\citenamefont
  {Zhachuk}, \citenamefont {Teys},\ and\ \citenamefont {Coutinho}}]{zha13}%
  \BibitemOpen
  \bibfield  {author} {\bibinfo {author} {\bibfnamefont {R.}~\bibnamefont
  {Zhachuk}}, \bibinfo {author} {\bibfnamefont {S.}~\bibnamefont {Teys}}, \
  and\ \bibinfo {author} {\bibfnamefont {J.}~\bibnamefont {Coutinho}},\ }\href
  {\doibase 10.1063/1.4808356} {\bibfield  {journal} {\bibinfo  {journal} {J.
  Chem. Phys.}\ }\textbf {\bibinfo {volume} {138}},\ \bibinfo {pages} {224702}
  (\bibinfo {year} {2013})}\BibitemShut {NoStop}%
\bibitem [{\citenamefont {Voigtl{\"a}nder}(2001)}]{voi01}%
  \BibitemOpen
  \bibfield  {author} {\bibinfo {author} {\bibfnamefont {B.}~\bibnamefont
  {Voigtl{\"a}nder}},\ }\href {\doibase 10.1016/S0167-5729(01)00012-7}
  {\bibfield  {journal} {\bibinfo  {journal} {Surf. Sci. Rep.}\ }\textbf
  {\bibinfo {volume} {43}},\ \bibinfo {pages} {127} (\bibinfo {year}
  {2001})}\BibitemShut {NoStop}%
\bibitem [{\citenamefont {Liu}\ \emph {et~al.}(2012)\citenamefont {Liu},
  \citenamefont {Camacho-Aguilera}, \citenamefont {Bessette}, \citenamefont
  {Sun}, \citenamefont {Wang}, \citenamefont {Cai}, \citenamefont {Kimerling},\
  and\ \citenamefont {Michel}}]{liu12}%
  \BibitemOpen
  \bibfield  {author} {\bibinfo {author} {\bibfnamefont {J.}~\bibnamefont
  {Liu}}, \bibinfo {author} {\bibfnamefont {R.}~\bibnamefont
  {Camacho-Aguilera}}, \bibinfo {author} {\bibfnamefont {J.~T.}\ \bibnamefont
  {Bessette}}, \bibinfo {author} {\bibfnamefont {X.}~\bibnamefont {Sun}},
  \bibinfo {author} {\bibfnamefont {X.}~\bibnamefont {Wang}}, \bibinfo {author}
  {\bibfnamefont {Y.}~\bibnamefont {Cai}}, \bibinfo {author} {\bibfnamefont
  {L.~C.}\ \bibnamefont {Kimerling}}, \ and\ \bibinfo {author} {\bibfnamefont
  {J.}~\bibnamefont {Michel}},\ }\href {\doibase 10.1016/j.tsf.2011.10.121}
  {\bibfield  {journal} {\bibinfo  {journal} {Thin Solid Films}\ }\textbf
  {\bibinfo {volume} {520}},\ \bibinfo {pages} {3354} (\bibinfo {year}
  {2012})}\BibitemShut {NoStop}%
\bibitem [{\citenamefont {Lozovoy}\ \emph {et~al.}(2014)\citenamefont
  {Lozovoy}, \citenamefont {Voytsekhovskiy}, \citenamefont {Kokhanenko},
  \citenamefont {Satdarov}, \citenamefont {Pchelyakov},\ and\ \citenamefont
  {Nikiforov}}]{loz14}%
  \BibitemOpen
  \bibfield  {author} {\bibinfo {author} {\bibfnamefont {K.~A.}\ \bibnamefont
  {Lozovoy}}, \bibinfo {author} {\bibfnamefont {A.~V.}\ \bibnamefont
  {Voytsekhovskiy}}, \bibinfo {author} {\bibfnamefont {A.~P.}\ \bibnamefont
  {Kokhanenko}}, \bibinfo {author} {\bibfnamefont {V.~G.}\ \bibnamefont
  {Satdarov}}, \bibinfo {author} {\bibfnamefont {O.~P.}\ \bibnamefont
  {Pchelyakov}}, \ and\ \bibinfo {author} {\bibfnamefont {A.~I.}\ \bibnamefont
  {Nikiforov}},\ }\href {\doibase 10.2478/s11772-014-0189-8} {\bibfield
  {journal} {\bibinfo  {journal} {Opto-Electron. Rev.}\ }\textbf {\bibinfo
  {volume} {22}},\ \bibinfo {pages} {171} (\bibinfo {year} {2014})}\BibitemShut
  {NoStop}%
\bibitem [{\citenamefont {Vanderbilt}(1987)}]{van87}%
  \BibitemOpen
  \bibfield  {author} {\bibinfo {author} {\bibfnamefont {D.}~\bibnamefont
  {Vanderbilt}},\ }\href {\doibase 10.1103/PhysRevB.36.6209} {\bibfield
  {journal} {\bibinfo  {journal} {Phys. Rev. B}\ }\textbf {\bibinfo {volume}
  {36}},\ \bibinfo {pages} {6209} (\bibinfo {year} {1987})}\BibitemShut
  {NoStop}%
\bibitem [{\citenamefont {J.~L.~Mercer}\ and\ \citenamefont
  {Chou}(1993)}]{mer93}%
  \BibitemOpen
  \bibfield  {author} {\bibinfo {author} {\bibfnamefont {J.}~\bibnamefont
  {J.~L.~Mercer}}\ and\ \bibinfo {author} {\bibfnamefont {M.~Y.}\ \bibnamefont
  {Chou}},\ }\href {\doibase 10.1103/PhysRevB.48.5374} {\bibfield  {journal}
  {\bibinfo  {journal} {Phys. Rev. B}\ }\textbf {\bibinfo {volume} {48}},\
  \bibinfo {pages} {5374} (\bibinfo {year} {1993})}\BibitemShut {NoStop}%
\bibitem [{\citenamefont {K{\"o}hler}\ \emph {et~al.}(1991)\citenamefont
  {K{\"o}hler}, \citenamefont {Jusko}, \citenamefont {Pietsch}, \citenamefont
  {M{\"u}ller},\ and\ \citenamefont {Henzler}}]{koh91}%
  \BibitemOpen
  \bibfield  {author} {\bibinfo {author} {\bibfnamefont {U.}~\bibnamefont
  {K{\"o}hler}}, \bibinfo {author} {\bibfnamefont {O.}~\bibnamefont {Jusko}},
  \bibinfo {author} {\bibfnamefont {G.}~\bibnamefont {Pietsch}}, \bibinfo
  {author} {\bibfnamefont {B.}~\bibnamefont {M{\"u}ller}}, \ and\ \bibinfo
  {author} {\bibfnamefont {M.}~\bibnamefont {Henzler}},\ }\href {\doibase
  10.1016/0039-6028(91)91178-Z} {\bibfield  {journal} {\bibinfo  {journal}
  {Surf. Sci.}\ }\textbf {\bibinfo {volume} {248}},\ \bibinfo {pages} {321}
  (\bibinfo {year} {1991})}\BibitemShut {NoStop}%
\bibitem [{\citenamefont {Koike}\ \emph {et~al.}(1997)\citenamefont {Koike},
  \citenamefont {Einaga}, \citenamefont {Hirayama},\ and\ \citenamefont
  {Takayanagi}}]{koi97}%
  \BibitemOpen
  \bibfield  {author} {\bibinfo {author} {\bibfnamefont {M.}~\bibnamefont
  {Koike}}, \bibinfo {author} {\bibfnamefont {Y.}~\bibnamefont {Einaga}},
  \bibinfo {author} {\bibfnamefont {H.}~\bibnamefont {Hirayama}}, \ and\
  \bibinfo {author} {\bibfnamefont {K.}~\bibnamefont {Takayanagi}},\ }\href
  {\doibase 10.1103/PhysRevB.55.15444} {\bibfield  {journal} {\bibinfo
  {journal} {Phys. Rev. B}\ }\textbf {\bibinfo {volume} {55}},\ \bibinfo
  {pages} {15444} (\bibinfo {year} {1997})}\BibitemShut {NoStop}%
\bibitem [{\citenamefont {Becker}\ \emph {et~al.}(1989)\citenamefont {Becker},
  \citenamefont {Swartzentruber}, \citenamefont {Vickers},\ and\ \citenamefont
  {Klitsner}}]{bec89}%
  \BibitemOpen
  \bibfield  {author} {\bibinfo {author} {\bibfnamefont {R.~S.}\ \bibnamefont
  {Becker}}, \bibinfo {author} {\bibfnamefont {B.~S.}\ \bibnamefont
  {Swartzentruber}}, \bibinfo {author} {\bibfnamefont {J.~S.}\ \bibnamefont
  {Vickers}}, \ and\ \bibinfo {author} {\bibfnamefont {T.}~\bibnamefont
  {Klitsner}},\ }\href {\doibase 10.1103/PhysRevB.39.1633} {\bibfield
  {journal} {\bibinfo  {journal} {Phys. Rev. B}\ }\textbf {\bibinfo {volume}
  {39}},\ \bibinfo {pages} {1633} (\bibinfo {year} {1989})}\BibitemShut
  {NoStop}%
\bibitem [{\citenamefont {Kim}\ \emph {et~al.}(2010)\citenamefont {Kim},
  \citenamefont {Oh}, \citenamefont {Baik}, \citenamefont {Jeon}, \citenamefont
  {Song}, \citenamefont {Nam}, \citenamefont {Woo}, \citenamefont {Park},\ and\
  \citenamefont {Ahn}}]{kim10}%
  \BibitemOpen
  \bibfield  {author} {\bibinfo {author} {\bibfnamefont {M.~K.}\ \bibnamefont
  {Kim}}, \bibinfo {author} {\bibfnamefont {D.-H.}\ \bibnamefont {Oh}},
  \bibinfo {author} {\bibfnamefont {J.}~\bibnamefont {Baik}}, \bibinfo {author}
  {\bibfnamefont {C.}~\bibnamefont {Jeon}}, \bibinfo {author} {\bibfnamefont
  {I.}~\bibnamefont {Song}}, \bibinfo {author} {\bibfnamefont {J.~H.}\
  \bibnamefont {Nam}}, \bibinfo {author} {\bibfnamefont {S.~H.}\ \bibnamefont
  {Woo}}, \bibinfo {author} {\bibfnamefont {C.-Y.}\ \bibnamefont {Park}}, \
  and\ \bibinfo {author} {\bibfnamefont {J.~R.}\ \bibnamefont {Ahn}},\ }\href
  {\doibase 10.1103/PhysRevB.81.085312} {\bibfield  {journal} {\bibinfo
  {journal} {Phys. Rev. B}\ }\textbf {\bibinfo {volume} {81}},\ \bibinfo
  {pages} {085312} (\bibinfo {year} {2010})}\BibitemShut {NoStop}%
\bibitem [{\citenamefont {Takayanagi}\ \emph {et~al.}(1985)\citenamefont
  {Takayanagi}, \citenamefont {Tanishiro}, \citenamefont {Takahashi},\ and\
  \citenamefont {Takahashi}}]{tak85}%
  \BibitemOpen
  \bibfield  {author} {\bibinfo {author} {\bibfnamefont {K.}~\bibnamefont
  {Takayanagi}}, \bibinfo {author} {\bibfnamefont {Y.}~\bibnamefont
  {Tanishiro}}, \bibinfo {author} {\bibfnamefont {S.}~\bibnamefont
  {Takahashi}}, \ and\ \bibinfo {author} {\bibfnamefont {M.}~\bibnamefont
  {Takahashi}},\ }\href {\doibase 10.1016/0039-6028(85)90753-8} {\bibfield
  {journal} {\bibinfo  {journal} {Surf. Sci.}\ }\textbf {\bibinfo {volume}
  {164}},\ \bibinfo {pages} {367} (\bibinfo {year} {1985})}\BibitemShut
  {NoStop}%
\bibitem [{\citenamefont {Soler}\ \emph {et~al.}(2002)\citenamefont {Soler},
  \citenamefont {Artacho}, \citenamefont {Gale}, \citenamefont {Garc{\'\i}a},
  \citenamefont {Junquera}, \citenamefont {Ordej{\'o}n},\ and\ \citenamefont
  {S{\'a}nchez-Portal}}]{sol02}%
  \BibitemOpen
  \bibfield  {author} {\bibinfo {author} {\bibfnamefont {J.~M.}\ \bibnamefont
  {Soler}}, \bibinfo {author} {\bibfnamefont {E.}~\bibnamefont {Artacho}},
  \bibinfo {author} {\bibfnamefont {J.~D.}\ \bibnamefont {Gale}}, \bibinfo
  {author} {\bibfnamefont {A.}~\bibnamefont {Garc{\'\i}a}}, \bibinfo {author}
  {\bibfnamefont {J.}~\bibnamefont {Junquera}}, \bibinfo {author}
  {\bibfnamefont {P.}~\bibnamefont {Ordej{\'o}n}}, \ and\ \bibinfo {author}
  {\bibfnamefont {D.}~\bibnamefont {S{\'a}nchez-Portal}},\ }\href {\doibase
  10.1088/0953-8984/14/11/302} {\bibfield  {journal} {\bibinfo  {journal} {J.
  Phys.: Condens. Matter}\ }\textbf {\bibinfo {volume} {14}},\ \bibinfo {pages}
  {2745} (\bibinfo {year} {2002})}\BibitemShut {NoStop}%
\bibitem [{\citenamefont {Perdew}\ and\ \citenamefont {Zunger}(1981)}]{per81}%
  \BibitemOpen
  \bibfield  {author} {\bibinfo {author} {\bibfnamefont {J.~P.}\ \bibnamefont
  {Perdew}}\ and\ \bibinfo {author} {\bibfnamefont {A.}~\bibnamefont
  {Zunger}},\ }\href {\doibase 10.1103/PhysRevB.23.5048} {\bibfield  {journal}
  {\bibinfo  {journal} {Phys. Rev. B}\ }\textbf {\bibinfo {volume} {23}},\
  \bibinfo {pages} {5048} (\bibinfo {year} {1981})}\BibitemShut {NoStop}%
\bibitem [{\citenamefont {Perdew}\ \emph {et~al.}(1996)\citenamefont {Perdew},
  \citenamefont {Burke},\ and\ \citenamefont {Ernzerhof}}]{per96}%
  \BibitemOpen
  \bibfield  {author} {\bibinfo {author} {\bibfnamefont {J.~P.}\ \bibnamefont
  {Perdew}}, \bibinfo {author} {\bibfnamefont {K.}~\bibnamefont {Burke}}, \
  and\ \bibinfo {author} {\bibfnamefont {M.}~\bibnamefont {Ernzerhof}},\
  }\href@noop {} {\bibfield  {journal} {\bibinfo  {journal} {Phys. Rev. Lett.}\
  }\textbf {\bibinfo {volume} {77}},\ \bibinfo {pages} {3865} (\bibinfo {year}
  {1996})}\BibitemShut {NoStop}%
\bibitem [{\citenamefont {Monkhorst}\ and\ \citenamefont {Pack}(1976)}]{mon76}%
  \BibitemOpen
  \bibfield  {author} {\bibinfo {author} {\bibfnamefont {H.~J.}\ \bibnamefont
  {Monkhorst}}\ and\ \bibinfo {author} {\bibfnamefont {J.~D.}\ \bibnamefont
  {Pack}},\ }\href {\doibase 10.1103/PhysRevB.13.5188} {\bibfield  {journal}
  {\bibinfo  {journal} {Phys. Rev. B}\ }\textbf {\bibinfo {volume} {13}},\
  \bibinfo {pages} {5188} (\bibinfo {year} {1976})}\BibitemShut {NoStop}%
\bibitem [{\citenamefont {Troullier}\ and\ \citenamefont
  {Martins}(1991)}]{tro91}%
  \BibitemOpen
  \bibfield  {author} {\bibinfo {author} {\bibfnamefont {N.}~\bibnamefont
  {Troullier}}\ and\ \bibinfo {author} {\bibfnamefont {J.~L.}\ \bibnamefont
  {Martins}},\ }\href {\doibase 10.1103/PhysRevB.43.1993} {\bibfield  {journal}
  {\bibinfo  {journal} {Phys. Rev. B}\ }\textbf {\bibinfo {volume} {43}},\
  \bibinfo {pages} {1993} (\bibinfo {year} {1991})}\BibitemShut {NoStop}%
\bibitem [{\citenamefont {Qin}\ \emph {et~al.}(2007)\citenamefont {Qin},
  \citenamefont {Shi}, \citenamefont {Ma}, \citenamefont {Gao}, \citenamefont
  {Rao}, \citenamefont {Wang},\ and\ \citenamefont {Pantelides}}]{qin07}%
  \BibitemOpen
  \bibfield  {author} {\bibinfo {author} {\bibfnamefont {Z.~H.}\ \bibnamefont
  {Qin}}, \bibinfo {author} {\bibfnamefont {D.~X.}\ \bibnamefont {Shi}},
  \bibinfo {author} {\bibfnamefont {H.~F.}\ \bibnamefont {Ma}}, \bibinfo
  {author} {\bibfnamefont {H.-J.}\ \bibnamefont {Gao}}, \bibinfo {author}
  {\bibfnamefont {A.~S.}\ \bibnamefont {Rao}}, \bibinfo {author} {\bibfnamefont
  {S.}~\bibnamefont {Wang}}, \ and\ \bibinfo {author} {\bibfnamefont {S.~T.}\
  \bibnamefont {Pantelides}},\ }\href {\doibase 10.1103/PhysRevB.75.085313}
  {\bibfield  {journal} {\bibinfo  {journal} {Phys. Rev. B}\ }\textbf {\bibinfo
  {volume} {75}},\ \bibinfo {pages} {085313} (\bibinfo {year}
  {2007})}\BibitemShut {NoStop}%
\bibitem [{\citenamefont {Tersoff}\ and\ \citenamefont {Hamann}(1985)}]{ter85}%
  \BibitemOpen
  \bibfield  {author} {\bibinfo {author} {\bibfnamefont {J.}~\bibnamefont
  {Tersoff}}\ and\ \bibinfo {author} {\bibfnamefont {D.~R.}\ \bibnamefont
  {Hamann}},\ }\href {\doibase 10.1103/PhysRevB.31.805} {\bibfield  {journal}
  {\bibinfo  {journal} {Phys. Rev. B}\ }\textbf {\bibinfo {volume} {31}},\
  \bibinfo {pages} {805} (\bibinfo {year} {1985})}\BibitemShut {NoStop}%
\bibitem [{\citenamefont {Horcas}\ \emph {et~al.}(2007)\citenamefont {Horcas},
  \citenamefont {Fern{\'a}ndez}, \citenamefont {G{\'o}mez-Rodr{\'i}guez},
  \citenamefont {Colchero}, \citenamefont {G{\'o}mez-Herrero},\ and\
  \citenamefont {Baro}}]{hor07}%
  \BibitemOpen
  \bibfield  {author} {\bibinfo {author} {\bibfnamefont {I.}~\bibnamefont
  {Horcas}}, \bibinfo {author} {\bibfnamefont {R.}~\bibnamefont
  {Fern{\'a}ndez}}, \bibinfo {author} {\bibfnamefont {J.~M.}\ \bibnamefont
  {G{\'o}mez-Rodr{\'i}guez}}, \bibinfo {author} {\bibfnamefont
  {J.}~\bibnamefont {Colchero}}, \bibinfo {author} {\bibfnamefont
  {J.}~\bibnamefont {G{\'o}mez-Herrero}}, \ and\ \bibinfo {author}
  {\bibfnamefont {A.~M.}\ \bibnamefont {Baro}},\ }\href {\doibase
  10.1063/1.2432410} {\bibfield  {journal} {\bibinfo  {journal} {Rev. Sci.
  Instrum.}\ }\textbf {\bibinfo {volume} {78}},\ \bibinfo {pages} {013705}
  (\bibinfo {year} {2007})}\BibitemShut {NoStop}%
\bibitem [{\citenamefont {Razado-Colambo}\ \emph {et~al.}(2009)\citenamefont
  {Razado-Colambo}, \citenamefont {He}, \citenamefont {Zhang}, \citenamefont
  {Hansson},\ and\ \citenamefont {Uhrberg}}]{raz09}%
  \BibitemOpen
  \bibfield  {author} {\bibinfo {author} {\bibfnamefont {I.}~\bibnamefont
  {Razado-Colambo}}, \bibinfo {author} {\bibfnamefont {J.}~\bibnamefont {He}},
  \bibinfo {author} {\bibfnamefont {H.~M.}\ \bibnamefont {Zhang}}, \bibinfo
  {author} {\bibfnamefont {G.~V.}\ \bibnamefont {Hansson}}, \ and\ \bibinfo
  {author} {\bibfnamefont {R.~I.~G.}\ \bibnamefont {Uhrberg}},\ }\href
  {\doibase 10.1103/PhysRevB.79.205410} {\bibfield  {journal} {\bibinfo
  {journal} {Phys. Rev. B}\ }\textbf {\bibinfo {volume} {79}},\ \bibinfo
  {pages} {205410} (\bibinfo {year} {2009})}\BibitemShut {NoStop}%
\bibitem [{\citenamefont {Ichikawa}\ and\ \citenamefont {Ino}(1984)}]{ich84}%
  \BibitemOpen
  \bibfield  {author} {\bibinfo {author} {\bibfnamefont {T.}~\bibnamefont
  {Ichikawa}}\ and\ \bibinfo {author} {\bibfnamefont {S.}~\bibnamefont {Ino}},\
  }\href {\doibase 10.1016/0039-6028(84)90611-3} {\bibfield  {journal}
  {\bibinfo  {journal} {Surf. Sci.}\ }\textbf {\bibinfo {volume} {136}}
  (\bibinfo {year} {1984}),\ 10.1016/0039-6028(84)90611-3}\BibitemShut
  {NoStop}%
\bibitem [{\citenamefont {Myslive{\v c}ek}\ \emph {et~al.}(2010)\citenamefont
  {Myslive{\v c}ek}, \citenamefont {Dvo{\v r}{\' a}k}, \citenamefont {Str{\'
  o}{\. z}ecka},\ and\ \citenamefont {Voigtl{\"a}nder}}]{mys10}%
  \BibitemOpen
  \bibfield  {author} {\bibinfo {author} {\bibfnamefont {J.}~\bibnamefont
  {Myslive{\v c}ek}}, \bibinfo {author} {\bibfnamefont {F.}~\bibnamefont
  {Dvo{\v r}{\' a}k}}, \bibinfo {author} {\bibfnamefont {A.}~\bibnamefont
  {Str{\' o}{\. z}ecka}}, \ and\ \bibinfo {author} {\bibfnamefont
  {B.}~\bibnamefont {Voigtl{\"a}nder}},\ }\href {\doibase
  10.1103/PhysRevB.81.245427} {\bibfield  {journal} {\bibinfo  {journal} {Phys.
  Rev. B}\ }\textbf {\bibinfo {volume} {81}},\ \bibinfo {pages} {245427}
  (\bibinfo {year} {2010})}\BibitemShut {NoStop}%
\bibitem [{\citenamefont {Zhachuk}\ and\ \citenamefont
  {Coutinho}(2011)}]{zha11}%
  \BibitemOpen
  \bibfield  {author} {\bibinfo {author} {\bibfnamefont {R.}~\bibnamefont
  {Zhachuk}}\ and\ \bibinfo {author} {\bibfnamefont {J.}~\bibnamefont
  {Coutinho}},\ }\href {\doibase 10.1103/PhysRevB.84.193405} {\bibfield
  {journal} {\bibinfo  {journal} {Phys. Rev. B}\ }\textbf {\bibinfo {volume}
  {84}},\ \bibinfo {pages} {193405} (\bibinfo {year} {2011})}\BibitemShut
  {NoStop}%
\bibitem [{\citenamefont {Henzler}(1982)}]{hen82}%
  \BibitemOpen
  \bibfield  {author} {\bibinfo {author} {\bibfnamefont {M.}~\bibnamefont
  {Henzler}},\ }\href {\doibase 10.1016/0378-5963(82)90092-7} {\bibfield
  {journal} {\bibinfo  {journal} {Applications of Surf. Sci.}\ }\textbf
  {\bibinfo {volume} {11-12}},\ \bibinfo {pages} {450} (\bibinfo {year}
  {1982})}\BibitemShut {NoStop}%
\bibitem [{\citenamefont {Dolbak}\ \emph {et~al.}(1989)\citenamefont {Dolbak},
  \citenamefont {Olshanetsky}, \citenamefont {Stenin},\ and\ \citenamefont
  {Teys}}]{dol89}%
  \BibitemOpen
  \bibfield  {author} {\bibinfo {author} {\bibfnamefont {A.~E.}\ \bibnamefont
  {Dolbak}}, \bibinfo {author} {\bibfnamefont {B.~Z.}\ \bibnamefont
  {Olshanetsky}}, \bibinfo {author} {\bibfnamefont {S.~I.}\ \bibnamefont
  {Stenin}}, \ and\ \bibinfo {author} {\bibfnamefont {S.~A.}\ \bibnamefont
  {Teys}},\ }\href {\doibase 10.1016/0039-6028(89)90619-5} {\bibfield
  {journal} {\bibinfo  {journal} {Surf. Sci.}\ }\textbf {\bibinfo {volume}
  {218}},\ \bibinfo {pages} {37} (\bibinfo {year} {1989})}\BibitemShut
  {NoStop}%
\bibitem [{\citenamefont {Neddermeyer}(1996)}]{ned96}%
  \BibitemOpen
  \bibfield  {author} {\bibinfo {author} {\bibfnamefont {H.}~\bibnamefont
  {Neddermeyer}},\ }\href {\doibase 10.1088/0034-4885/59/6/001} {\bibfield
  {journal} {\bibinfo  {journal} {Rep. Prog. Phys.}\ }\textbf {\bibinfo
  {volume} {59}},\ \bibinfo {pages} {701} (\bibinfo {year} {1996})}\BibitemShut
  {NoStop}%
\end{thebibliography}
%

\end{document}